\documentclass{aa}  
\usepackage{subfigure}
\usepackage{threeparttable}
\usepackage{parskip}
\usepackage{float}
\usepackage{hyperref}
\usepackage{graphicx}
\usepackage{color}
\usepackage{url}
\usepackage[normalem]{ulem}
\usepackage{txfonts}

\newcommand{\xmm}{{\it XMM-Newton\/}}

\def\be{\begin{equation}} 
\def\ee{\end{equation}}

\def\be{\begin{equation}} 
\def\ee{\end{equation}}

\usepackage{graphicx}
\usepackage{txfonts}
\begin{document} 

\title{Emission and absorption lines in the photospheric radius expansion bursts of 4U 1820--30}

   \subtitle{}

\author{Wenhui Yu\inst{1}
        \and
        Zhaosheng Li\inst{1}
        \thanks{Corresponding author}
        \and
        Yuanyue Pan\inst{1}
        \thanks{Corresponding author}
        \and
        Xuejuan Yang\inst{1}
        \and
         Yupeng Chen\inst{2}
        \and
         Shu Zhang\inst{2}
        \and
         Maurizio Falanga \inst{3,4}
         }
   \offprints{Z. Li}

   \institute{Key Laboratory of Stars and Interstellar Medium, Xiangtan University, Xiangtan 411105, Hunan, P.R. China\\ \email{lizhaosheng@xtu.edu.cn, panyy@xtu.edu.cn}
       \and
Key Laboratory of Particle Astrophysics, Institute of High Energy Physics, Chinese Academy of Sciences, 19B Yuquan Road, Beijing 100049, China
   \and
International Space Science Institute (ISSI), Hallerstrasse 6, 3012 Bern, Switzerland
           \and
Physikalisches Institut, University of Bern, Sidlerstrasse 5, 3012 Bern, Switzerland
}

   \date{Received XX; accepted XX}
\abstract
{We analyze the emission and absorption lines during photospheric radius expansion (PRE) X-ray bursts from the ultracompact binary 4U 1820--30, observed with the Neutron Star Interior Composition Explorer (NICER).  Using Monte Carlo simulations to estimate the significance, we identified a 1 kiloelectron volt (keV) emission line from 14 bursts, a 3 keV absorption line from 12 bursts, and a 1.6 keV absorption line from one burst.  By coadding the burst spectra at the maximum radius phase, we detected a 1.034 keV emission line with significance of $14.2\sigma$, along with absorption lines at 1.64 and 3 keV, with significances of $10.8\sigma$ and $11.7\sigma$, respectively. The observed energy shifts are consistent with the prediction from the burst-driven wind model, indicating that all three spectral features are produced by the PRE wind. An analysis of the ratios between the emission and absorption line energies suggests that the 1 keV feature is a superposition of several narrower Fe L-shell lines. To evaluate the scientific capabilities of the Hot Universe Baryon Surveyor (HUBS), we simulated mock observations of multiple narrow lines near 1 keV. These results demonstrate that HUBS is well suited for detailed studies of the 1 keV emission line during bursts, offering significant potential to advance our understanding of these phenomena.}

\keywords{Stars: neutron - X-rays: bursts – accretion, accretion disks – X-rays: binaries--X-rays: individuals: 4U 1820--30}
\titlerunning{4U 1820--30}
\authorrunning{Yu et al.}
   \maketitle

\section{Introduction}
 The 4U 1820--30 system is a persistent, ultracompact, X-ray binary (UCXB) with an orbital period of 11.4 minutes \citep{Stella1987}. Due to its short orbital period, the donor star is likely a white dwarf  with a mass of $\sim 0.06-0.07 ~M_{\odot}$ \citep{1981Paczynski, Nelson1986ApJ, Rappaport1987}.  In this system, a neutron star (NS) accretes helium-rich material from its companion, potentially triggering unstable thermonuclear burning, resulting in type I X-ray bursts \citep[or simply X-ray burst;][]{Lewin93,Strohmayer06,2008A&AFalanga,galloway2008thermonuclear,Galloway21}. These bursts typically release a total energy of $\sim 10^{39}$ erg over 10--100 s. Occasionally, bursts can exceed the Eddington limit, causing the NS surface layers to lift, resulting in photospheric radius expansion  \citep[PRE;][]{Lewin93}. In rare cases, the photosphere radius can expand by a factor of $>100$, described as a ``superexpansion'' burst \citep{int2010,Yuwh2024A&A}.

A strong PRE burst can drive the matter outflow from the NS surface, such as generating a super-Eddington wind \citep{Paczynski1986ApJ,Yu2018,Guichandut2021ApJ}. There is some evidence suggesting that the wind is polluted with heavy nuclear burning ashes \citep{int2010,Yu2018,Guichandut2021ApJ}, as type I X-ray bursts are known sites for the production of heavy element via the $rp$-process \citep{SCHATZ2003247}. Convection at the onset of the burst can mix heavy elements from the deeper burning layers, which are then ejected by the wind and exposed at the photosphere \citep{Weinberg2006,Yu2018}. In superexpansion bursts, the column density of ejected ashes is expected to be particularly high due to the energetic nature of these events. As a result, the heavy element ashes could be observed as spectral lines or absorption edges \citep{int2010}. For example, in two superexpansion bursts from 4U 0614+091 and 4U 1722--30 detected by RXTE, \citet{int2010} reported spectral edges in the range of 6--11 kiloelectron volts (keV). These edge energies can be associated with a redshifted hydrogen or helium-like Ni edge, suggesting the presence of heavy-element ashes in the wind. \citet{li2018ApJ} identified an absorption edge in a PRE burst from GRS 1747--312 during RXTE\ observations. The edge energy, remaining at $\approx 8~{\rm keV}$ during the cooling tail, was attributed to photoionization absorption edges of hydrogen-like nickel, redshifted to match the observed energy. Based on this finding, they estimated the NS mass, radius, and surface gravitational redshift factor.

In 2017, \citet{Keek2018} reported a strong PRE burst in 4U 1820--30 from NICER observations. In addition to a blackbody component, an extra optically thick Comptonization model has to be added to fit the time-resolved burst spectra.  Four of these bursts exhibited significant spectral features, including an emission line at 1 keV and absorption features at 1.7 keV and 3 keV \citep{Strohmayer2019}. They identified a relative spectral shift of approximately 1.046 in bursts with a photospheric radius of around 75 km, attributed to the combined effects of gravitational redshift and Doppler blueshift in a burst-driven wind.  However, from the calculations in \citet{Guichandut2021ApJ}, only a static envelope can be produced for such a small photospheric radius and the wind-induced blueshift is negligible.  More recently, \citet{Yuwh2024A&A} analyzed 15 type I X-ray bursts from 4U 1820--30 using NICER \citep[see also][]{Jaisawal24}. The time-resolved burst spectra revealed that all these bursts were PRE with photospheric radii ranging from 55 to $10^3$ km. Additionally, they provided direct evidence that the accretion disk was distorted by the burst radiation by fitting the burst spectra with a reflection model.

In this work, extending the analysis by \citet{Strohmayer2019} and \citet{Yuwh2024A&A},  we focus on the spectral lines of 15 X-ray bursts from 4U 1820--30 observed by NICER. This work is organized as follows. In Sect. \ref{Sec:observe}, we describe how we performed the spectral line fitting during the X-ray bursts. We discuss and summarize the results in Sects. \ref{Sec:discussion} and \ref{Sec:conclusion}, respectively.

\section{Observations and data reduction}
\label{Sec:observe} 
\begin{figure*}
\includegraphics[width=\hsize]{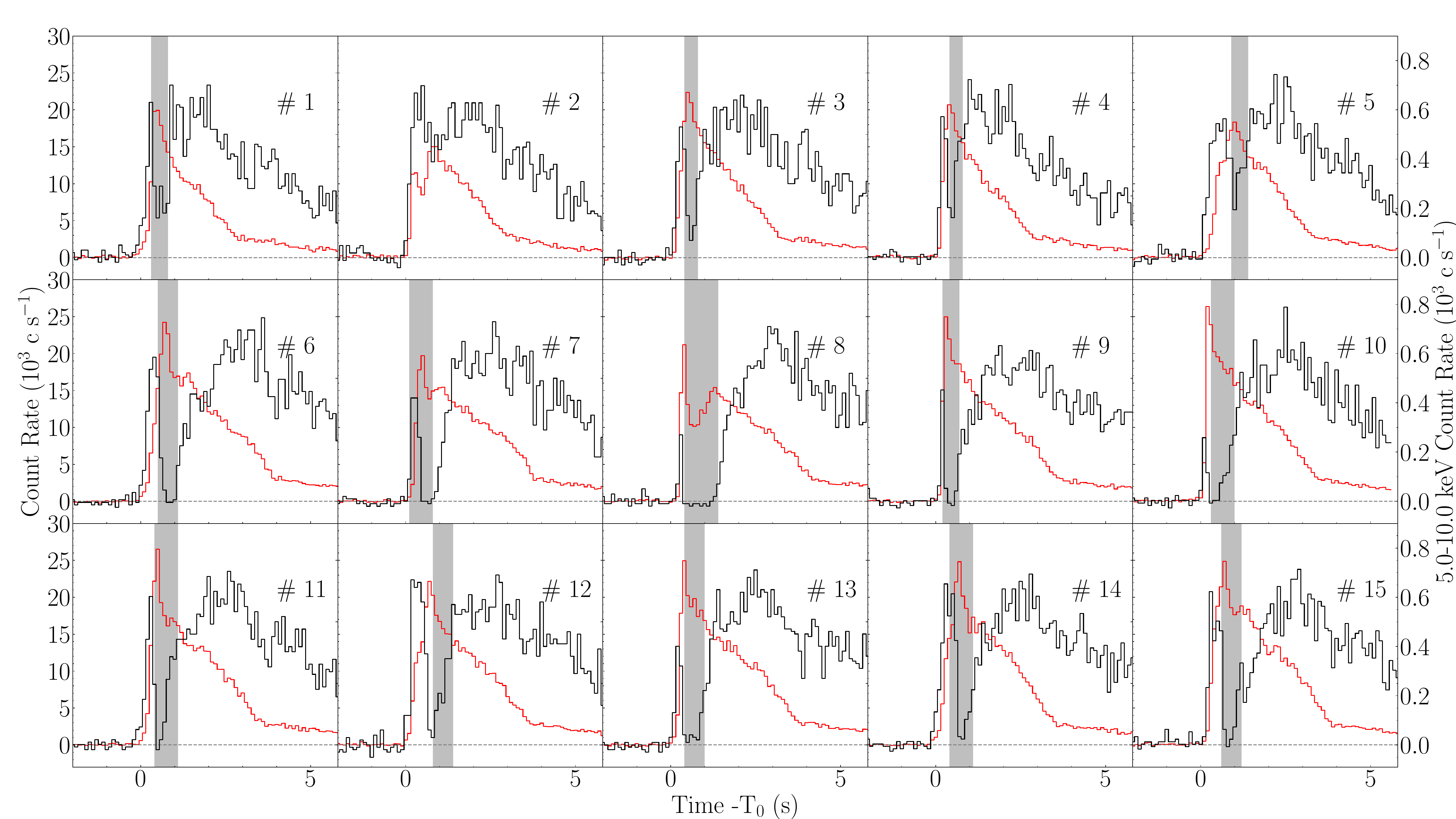}
   \caption{0.1 s binned light curves of 15 X-ray bursts from 4U 1820--30 observed by NICER in 0.5--10 keV are marked as red  lines.  The right axis shows the 3--10 keV light curves as black lines. The persistent emissions are regarded as background and subtracted. The gray shaded region mark the intervals used to extract PRE phase spectra. }
      \label{fig:lc}
\end{figure*}

NICER observed 4U 1820--30 from 2017 to present for a total exposure time of 812 ks, in which 15 type I X-ray bursts were detected with a short duration and a high peak count rate. We processed all archived NICER data by applying the standard filtering criteria using HEASOFT V6.33.2 and the NICER Data Analysis Software (NICERDAS) V2.0.7. The 0.1 s binned light curves were extracted using \texttt{nicerl3-lc} from the calibrated unfiltered (UFA) files and cleaned events files, see Fig. \ref{fig:lc}. We extracted the spectra, ancillary response files ($\texttt{ARFs}$), response matrix files ($\texttt{RMFs}$), and the 3c50 background spectra  \citep[][]{Remillard21} using \texttt{nicerl3-spect}. We performed the spectra analysis using Xspec v12.14.1 \citep{Arnaud96}. The errors of all parameters are quoted at the $1\sigma$ confidence level.

All time-resolved burst spectra can be fitted well by using the enhanced persistent emission model ($f_{\rm a}$ model), adding a reflection component from the surrounding accretion disk (disk reflection model), or using the double-blackbody model, as reported by \citet{Yuwh2024A&A}. All bursts showed the characteristics of PRE. However, the $f_a$ model failed to provide a physically motivated and self-consistent explanation for the observations. This study focuses on the spectral features of X-ray bursts from 4U 1820--30, as detected by NICER.  Due to its phenomenological simplicity, the double-blackbody model was adopted to describe the continuum spectra for subsequent analysis. For more details on continuum fitting, we refer to \citet{Yuwh2024A&A}.

\begin{figure}
\includegraphics[width=\hsize]{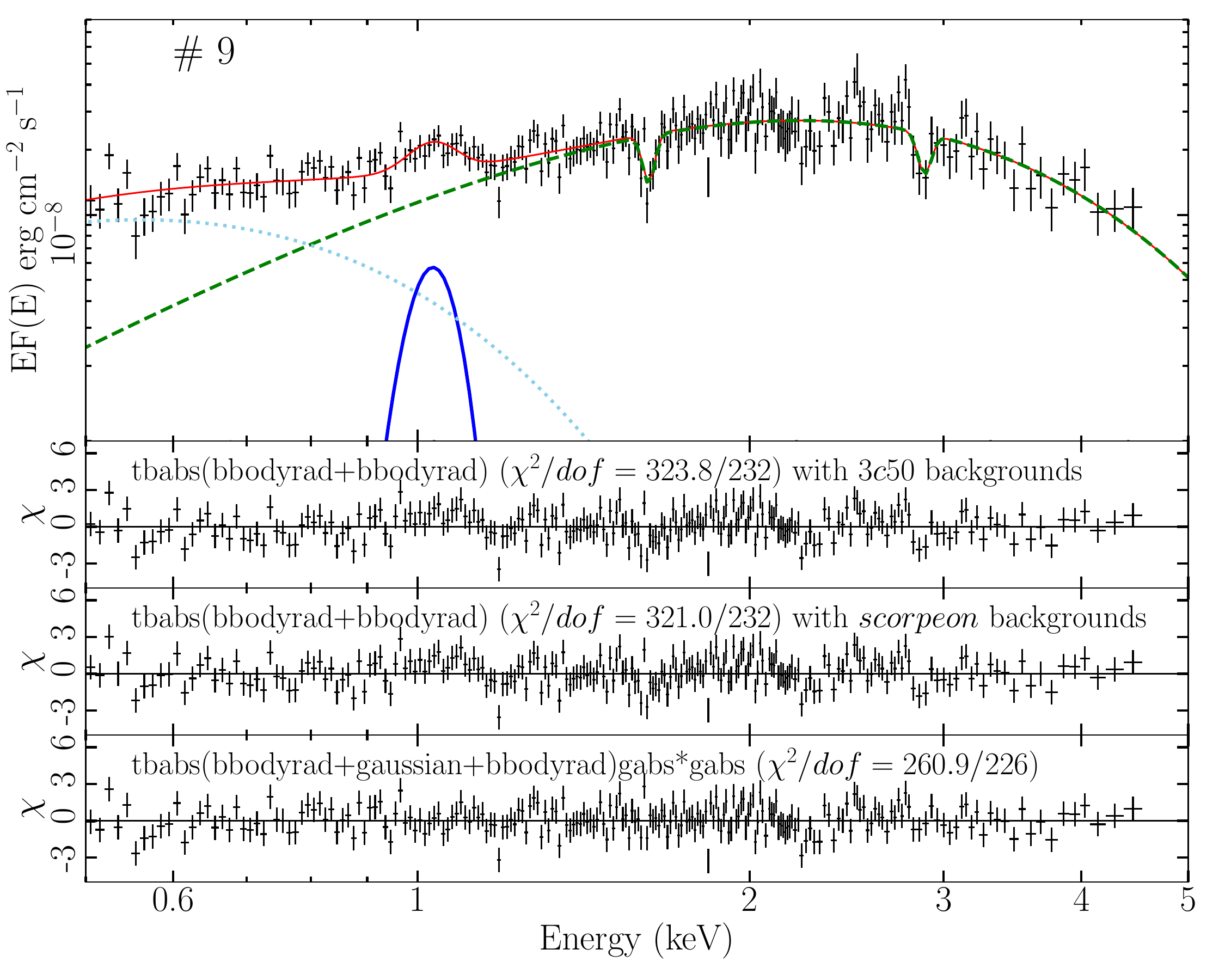}
      \label{fig:single}
         \caption{ PRE phase spectra and residuals for burst \#\ 9. {\it Top panel}:  PRE phase spectra from the burst \# 9. The best-fit model, ${\tt Tbabs} \times ({\tt bbodyrad + gaussion +bbodyrad })\times {\tt gabs} \times {\tt gabs}$, is represented by the  solid red line. The 1 keV emission line are clearly shown in the spectra and a pair of absorption lines are indicated near 1.6 and 3 keV.  The two middle panels are the residuals of the spectra fitted with 3c50 and SCORPEON backgrounds, respectively. The bottom panel shows the residuals of the best-fit model to the spectra. }
\end{figure}

\begin{figure}
\includegraphics[width=\hsize]{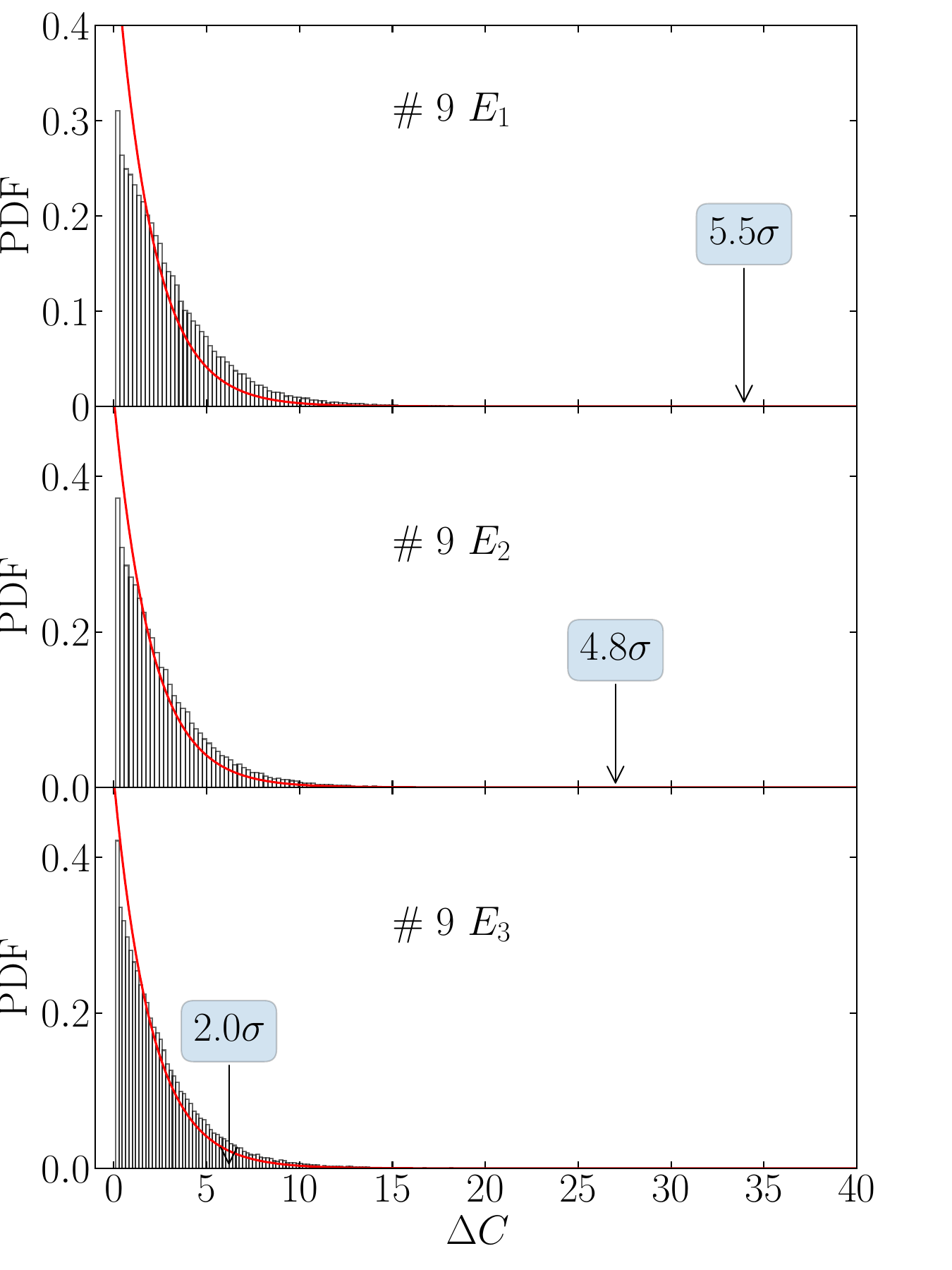}
   \caption{Distributions of the $\Delta C$ of $10^5$  simulations for the lines of burst $\#\ 9$. The distributions fitted by  $\chi^2$-distribution with two dofs (red dashed-dotted lines). From top to bottom panel, the arrow marks the significance for the 1 keV emission line, 1.6 and 3.0 keV absorption lines, respectively.}
      \label{fig:bur9_mc}
\end{figure}
 
\subsection{Single burst spectra} 
\label{Sec:single burst spectra} 
 \citet{Strohmayer2019} found that the spectral lines appeared only during the PRE phase. No spectral lines were present in the pre-burst persistent spectra and the initial and cooling phases of the bursts \citep{Yuwh2024A&A}. Therefore, we extracted spectra during the PRE phase for each burst based on the time-resolved spectroscopy. The spectra were extracted with the exposure time varying
from 0.3 to 2 s.  The gray area in Fig.~\ref{fig:lc} marks the intervals used to extract PRE phase spectra. The burst spectra were grouped using {\tt grappha} with a minimum count of 20 and regarded the persistent spectra as background and unchanged during bursts. The PRE spectra of most burst were dominated by the persistent emission above 5 keV, see the burst light curve in 5--10 keV after subtracting the persistent count rate in Fig.~\ref{fig:lc}.   To focus on searching for spectral features, we fitted each spectrum in 0.5--5 keV with the double-blackbody model, namely, \texttt{TBabs} $\times$ \texttt{(bbodyrad} + \texttt{bbodyrad)}. The absorption column density was fixed as the best fitting value from the persistent spectra, $N_{\rm H} = 1.93 \times 10^{21}~{\rm cm} ^{-2}$.

The double-blackbody model provides a reasonable qualitative fit to the continuum spectra, but there are substantial spectral features evident in the residual. For each X-ray burst, we selected a spectrum with significant emission and absorption lines to analyze the spectral lines of the burst.   In Fig.~\ref{fig:single}, as an example, we show the PRE phase spectrum extracted from burst \#~9, which has $\chi^{2}=315.82$ for 230 degrees of freedom (dofs) fitted by the double-blackbody model. The residuals showed apparent line features near 1.0 and 1.6 keV, and a possible 3 keV absorption line guided by \citet{Strohmayer2019}. Adding one Gaussian (\texttt{gauss}) and two absorption components (\texttt{gabs}) to describe these features; namely, the overall model is ${\tt Tbabs} \times ({\tt bbodyrad + gaussion +bbodyrad })\times {\tt gabs} \times {\tt gabs}$, which can obviously improve the fitting result. The $\chi^{2}$ is 256.51 for 224 dofs. The best-fit centroid energies and line width are $E_{1} = 1.028\pm{0.024}$ keV and width $\sim0.05$ keV, $E_{2} = 1.625\pm{0.021}$ keV and width $\sim0.02$ keV, and  $E_{3} = 2.894\pm{0.055} $ keV and width $\sim0.04$, respectively. 

To confirm the robustness of spectral features against background modeling uncertainties, we reprocessed the data using \texttt{nicerl3-spec} with the SCORPEON background model \footnote{\href{https://heasarc.gsfc.nasa.gov/docs/nicer/analysis_threads/scorpeon-xspec/}{https://heasarc.gsfc.nasa.gov/docs/nicer/analysis$\_$threads/scorpeon-xspec/}}. Using burst \# 9 as an example, we compared the spectral lines from different background models. The best-fit parameters of the 1 keV emission and 3 keV absorption lines were consistent within $1\sigma$ confidence level compared with the 3c50 background model (see Fig.~\ref{fig:single}).

\subsection{Spectra line significance assessment}
\label{Sec:significance} 

The significance of spectral lines was not reliably determined using goodness-of-fit or F-test methods, particularly for the emission and absorption lines  \citep{Protassov_2002,Wang2022}. The F-test tends to overestimate the significance of these features \citep{Protassov_2002,2024A&AG.L}. Therefore, we employed the Monte Carlo (MC) method to evaluate the significance of potential spectral lines  \citep{2004A&APorquet,Tombesi2010A&A,Gofford13}.  This approach has proven robust and has become the standard practice in the past decade \citep{Li_2022,Xu_2022,Parra2024A&A,zhang2024}

To illustrate the MC method, we used burst \#\ 9 as an example and analyzed three spectral features independently. The null hypothesis model was defined as the best-fit model for the burst spectrum without any spectral lines, described by \texttt{TBabs} $\times$ \texttt{(bbodyrad} + \texttt{bbodyrad)}. The observed spectrum was fitted using the null hypothesis model, and we obtained a baseline C-stat value $C_{\rm null}$ using C-statistics. Subsequently, we added a spectral line to the model, yielding an improved fit and a new C-stat value $C_{\rm line}$. The likelihood ratio, represented by the difference $\Delta C=C_{\rm null}-C_{\rm line}$ was then calculated to evaluate the statistical significance of the spectral feature \citep{cash1979Apj, zhang2024}, 
\begin{equation}
\begin{split}
    LR
    &=2\ln\frac{L_{\rm line}}{L_{\rm null}}\\
    &=-2\ln L_{\rm null}-(-2\ln L_{\rm line})\\
    &=C_{\rm null}-C_{\rm line}\\
    &=\Delta C \sim \chi^{2} (\mathrm{x, dof = 2}),
    \label{eq:delta_c}
\end{split}      
\end{equation}
where $L_{\rm null}$ and $L_{\rm line}$ represent the likelihoods for the null hypothesis model and the model with an additional line component, respectively. The $\Delta C_{\rm true}$ for three spectral features are calculated as, $\Delta C_{{\rm true},~E_{1}} = 33.93$, $\Delta C_{{\rm true},~E_{2}} = 27.01$, and $\Delta C_{{\rm true},~E_{3}} = 6.20$, respectively. To assess these features, we generate $10^5$ distributions of best-fitted parameters within the uncertainties using \texttt{simpars} command in Xspace. Each set of model parameters was loaded from the generated distribution, and a spectrum was simulated using the \texttt{fakeit} command in Xspec, preserving all observational parameters (exposure, response files, background) of the initial spectrum. The simulated spectra were first fitted with the double-blackbody model to obtain the baseline $C_{\rm null}$ values. Then, an emission or absorption line component was added to the model and the spectra were re-fitted to obtain $C_{\rm line}$. The resulting $\Delta C_{\rm{sim}}$ values were recorded for each simulated spectrum, forming a distribution for statistical analysis. This method provides a robust estimate of the significance of the spectral features.

For the $10^5$ simulations, the distributions of $\Delta C_{\text{sim}}$ were fitted to a $\chi^2$-distribution, and the $p$-values were inferred based on these fits. Figure~\ref{fig:bur9_mc} illustrates the distributions and fitting results for the three spectral features of burst \#\ 9. The results show that the simulated distribution is consistent with the $\chi^2$-distribution with two dofs. The two dofs of $\chi^2$ are derived from the two free parameters of gaussian or absorption given the fixed line width. The derived significances are, $\sigma_{\rm E1} = 5.5\sigma$, $\sigma_{\rm E2} = 4.8\sigma$, and $\sigma_{\rm E3} = 2.0\sigma$. 

\begin{figure*}

    \centering
        \includegraphics[width=\hsize]{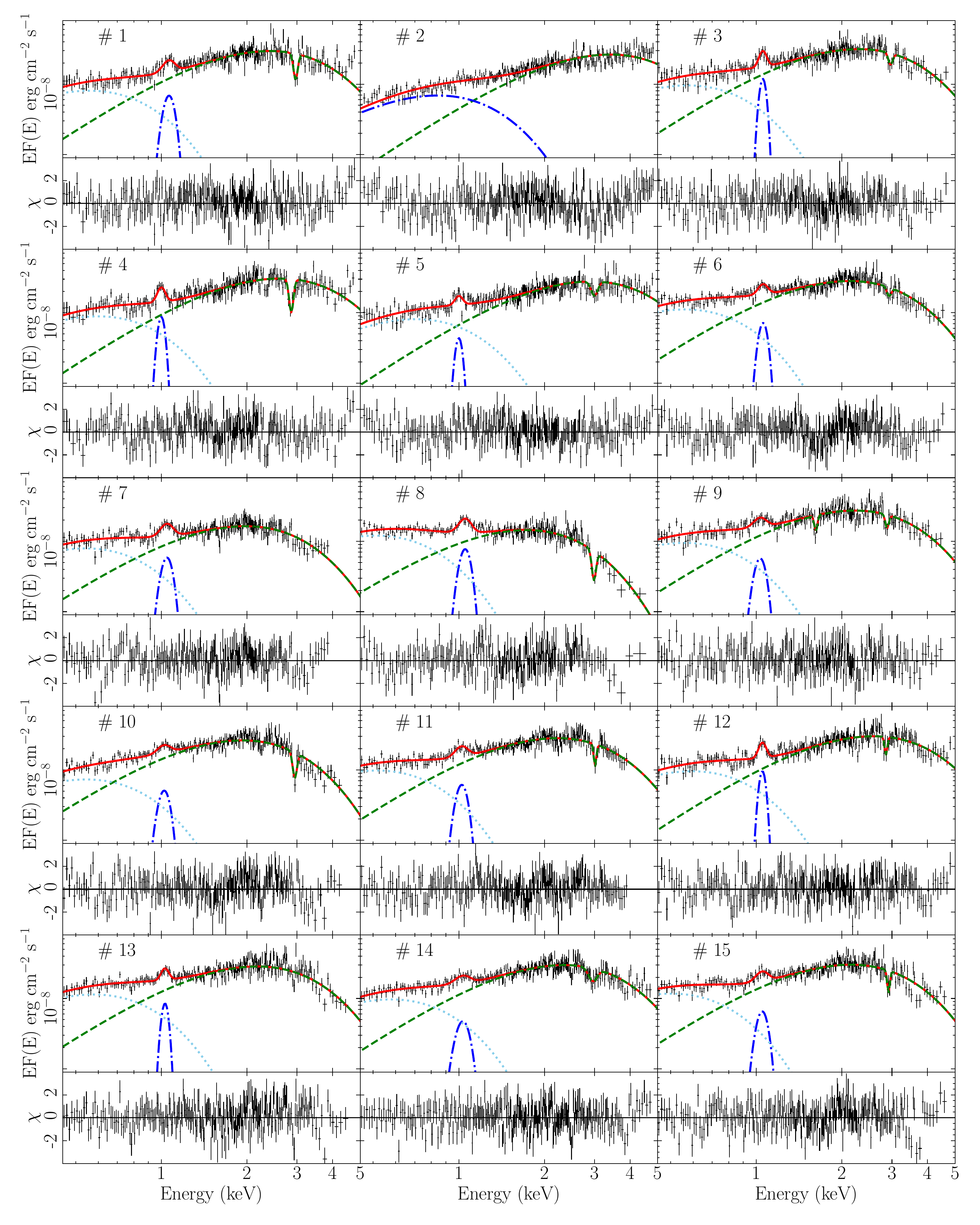}

\caption{Unabsorbed spectra extracted during the PRE phase for all bursts in 0.3--5 keV. The  solid red lines indicate the best-fitting model, including the continuum and spectral lines if necessary. The residuals of the best fit model to the data are plotted in bottom panels.}
      \label{fig:line}
\end{figure*}

The MC method was subsequently applied to all spectra to assess the significance of detected lines. Only spectral lines with a significance greater than $2\sigma\ (95\%)$ were considered detections and included in the further analyses (see Table~\ref{table:burst_PRE1}). For burst \# 2, no spectral lines were detected, consistent with the findings of \citet{Strohmayer2019}. 
In total, we identified the 1 keV emission line in 14 bursts, the 3 keV absorption line in 12 bursts, and the 1.6 keV absorption line in one burst (\#\ 9). The spectra and the best-fitted model are shown in Fig.~\ref{fig:line}. Table~\ref{table:burst_PRE1} summarizes the results of best-fit parameters of the emission and absorption lines. The strongest identified feature is the 1 keV emission line from burst $\#\ 8$ with a significance of $10.8\sigma$. All spectral features exhibit narrow line widths of approximately 0.02--0.08 keV; thus, the line width was fixed during fitting.
  
\begin{table*}
\begin{center} 
\caption{Gaussian line parameters.  \label{table:burst_PRE1}}
\resizebox{\linewidth}{!}{\begin{tabular}{cccccccccccccccccccc} 
 & \multicolumn{4}{c}{ Emission line, $E_1$}& &\multicolumn{4}{c}{ Absorption line, $E_2$}& &\multicolumn{4}{c}{Absorption line, $E_3$} \\
\cline{2-5} \cline{7-10} \cline{12-15} \\
{\centering  Burst  } &
{\centering  $E_1$} &
{\centering  Norm$_1^{\rm a}$} &
{\centering  Width$_1$}&
{\centering  $P_1^{\rm b}$}&
&
{\centering  $E_2$} &
{\centering  Norm$_2^{\rm a}$} &
{\centering  Width$_2$}&
{\centering  $P_2^{\rm b}$}&
&
{\centering  $E_3$} &
{\centering  Norm$_3^{\rm a}$} &
{\centering  Width$_3$}&
{\centering  $P_3^{\rm b}$}&
{\centering  $\chi^{2}(\mathrm{dof})$}&
\\
$\#$&$\mathrm{(keV)}$&&$\mathrm{(keV)}$&$(\sigma)$&&$\mathrm{(keV)}$&&$\mathrm{(keV)}$&$(\sigma)$&&$\mathrm{(keV)}$&&$\mathrm{(keV)}$&$(\sigma)$&\\ [0.01cm] \hline
1  & $1.052\pm{0.018}$ & $0.459\pm{0.111}$ & 0.05&5.2 &&                 &                 &    &    &&$2.957\pm{0.031}$&$0.093\pm{0.024}$&0.05&5.3&300(237)\\
2  & -                 & -                 & -   &  -  &&                 &                 &    &    &&       -         &     -          &  - &-&276(258)\\
3  & $1.051\pm{0.017}$ & $0.647\pm{0.199}$ & 0.04&5.9 &&                 &                 &    &    &&$2.977\pm{0.036}$&$0.098\pm{0.045}$&0.05&3.3&197(199)\\
4  & $0.997\pm{0.014}$ & $0.422\pm{0.122}$ & 0.03&4.3 &&                 &                 &    &    &&$2.860\pm{0.028}$&$0.127\pm{0.052}$&0.05&5.9&246(225)\\
5  & $0.996\pm{0.024}$ & $0.244\pm{0.105}$ & 0.03&3.2 &&                 &                 &    &    &&$2.994\pm{0.077}$&$0.075\pm{0.032}$&0.08&4.0&256(238)\\
6  & $1.046\pm{0.021}$ & $0.405\pm{0.106}$ & 0.04&5.5 &&                 &                 &    &    &&$2.932\pm{0.038}$&$0.066\pm{0.031}$&0.05&3.1&279(236)\\
7  & $1.038\pm{0.017}$ & $0.398\pm{0.091}$ & 0.05&6.1 &&                 &                 &    &    &&                 &                 &    &    &272(231)\\
8  & $1.052\pm{0.012}$ & $0.456\pm{0.070}$ & 0.05&10.8&&                 &                 &    &    &&$2.967\pm{0.044}$&$0.069\pm{0.033}$&0.05&3.2&280(235)\\
9  & $1.028\pm{0.024}$ & $0.484\pm{0.138}$ & 0.05&5.5 &&$1.625\pm{0.021}$&$0.032\pm{0.011}$&0.02&4.8 &&$2.894\pm{0.055}$&$0.040\pm{0.011}$&0.04&2.0&256(224)\\
10 & $1.034\pm{0.017}$ & $0.406\pm{0.105}$ & 0.05&6.5 &&                 &                 &    &    &&$2.963\pm{0.024}$&$0.083\pm{0.028}$&0.05&4.3&310(253)\\
11 & $1.018\pm{0.022}$ & $0.392\pm{0.102}$ & 0.05&7.5 &&                 &                 &    &    &&$3.014\pm{0.051}$&$0.039\pm{0.024}$&0.03&2.0&245(260)\\
12 & $1.043\pm{0.013}$ & $0.374\pm{0.099}$ & 0.03&5.7 &&                 &                 &    &    &&$2.863\pm{0.057}$&$0.040\pm{0.022}$&0.03&2.1&256(251)\\
13 & $1.029\pm{0.018}$ & $0.327\pm{0.103}$ & 0.03&5.1 &&                 &                 &    &    &&                 &                 &    &    &280(249)\\
14 & $1.019\pm{0.014}$ & $0.427\pm{0.113}$ & 0.06&5.4 &&                 &                 &    &    &&$2.926\pm{0.023}$&$0.129\pm{0.028}$&0.08&6.2&245(252)\\
15 & $1.053\pm{0.019}$ & $0.456\pm{0.115}$ & 0.05&6.9 &&                 &                 &    &    &&$2.927\pm{0.027}$&$0.079\pm{0.033}$&0.03&4.4&259(244)\\
\hline
\end{tabular} }
\end{center}
$^{\rm a}$ The line normalization of \texttt{gauss} or \texttt{gabs} in units of $\mathrm{photons~cm^{-2}~s^{-1} }$.

$^{\rm b}$  The significance of the lines were estimated using the Monte Carlo simulations.
\end{table*}

\subsection{Coadded spectra}
\label{Sec:Coadded spectra} 
\begin{figure}
    \centering
        \includegraphics[width=9cm]{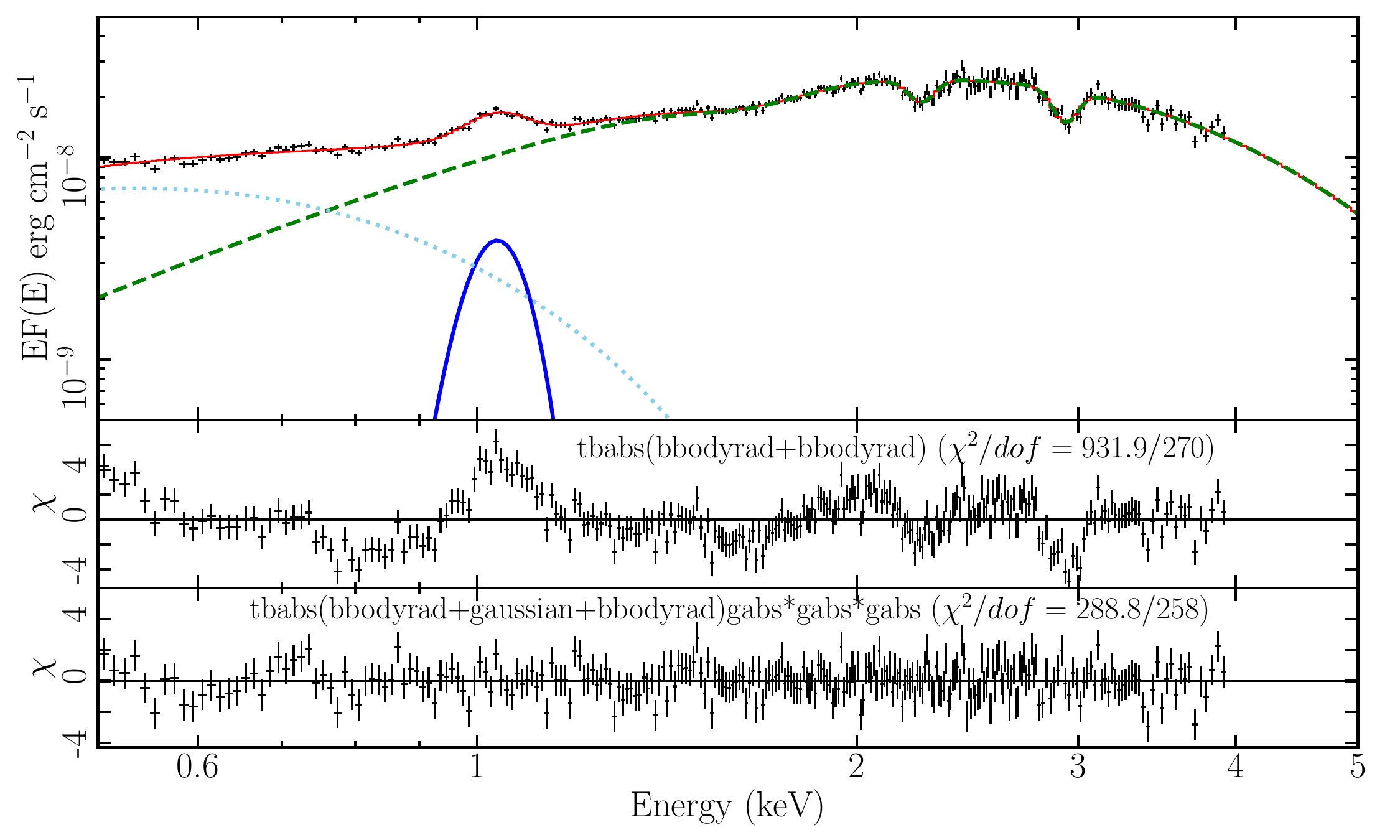}
\caption{Coadded spectroscopy of all bursts (except burst \# 2) at the maximum photospheric radius. The  solid red line is the best-fitting model including the continuum and the emission line, $E_{\rm 1}$, the absorption lines, $E_{\rm 2}$ and $E_{\rm 3}$, and the absorption lines at 2.2 keV (due to the reflectivity of gold M shell of the detector itself). The middle panel shows the residuals of the fitting of the model without lines. The residuals of the best fit model to the data are plotted in the bottom panel.}
      \label{fig:line_coad}
\end{figure}

\citet{Guichandut2021ApJ} presented the hydrodynamic simulation of super Eddington winds. Their interpretation states that the observed energy will close to the line energy at the rest frame where the photospheric layer expands to a large radius (more than 200 km). In our case, as reported by \citet{Yuwh2024A&A}, we found that the maximum radii of 14 bursts exceeded 100 km. However, the photospheric expends so fast during the PRE phase that we cannot obtain a significant line around the maximum radius in one burst. To address this, we stacked all burst spectra for a 0.1 s interval at the maximum radius, excluding burst \# 2, to enhance the signal-to-noise ratio (S/N). A double-blackbody model was applied to fit the coadded spectra. Figure~\ref{fig:line_coad} shows the stacked spectrum from the strong photospheric expansion phase, with the red curve representing the best-fitting model.   We found that the hotter blackbody component with a temperature $kT_\mathrm{bb1}= 0.58\pm{0.01} $ keV and a radius of $R_\mathrm {bb1}=117\pm{26}$ km, while the cooler blackbody component has a temperature of $kT_{\mathrm {bb2}}= 0.17\pm{0.01}$ keV and a blackbody radius of $R_\mathrm {bb2}=1146\pm{335}$ km.

\begin{figure}

    \centering
        \includegraphics[width=\hsize]{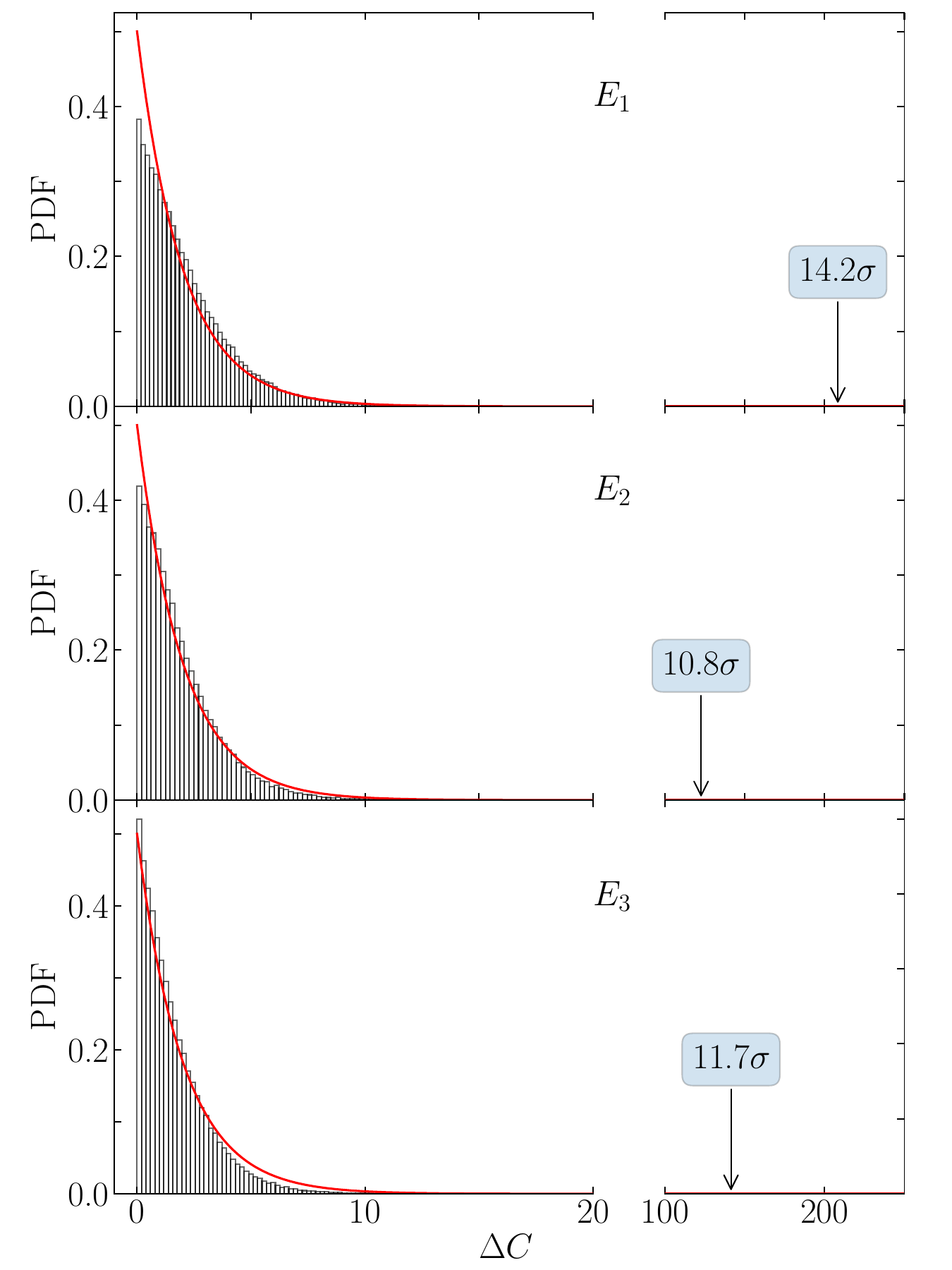}

\caption{Distributions of the $\Delta C$  for $10^5$  simulations for the coadded spectra lines. From top to bottom: Distributions of the  simulations for 1 keV emission line ($E_1$), 1.6 keV absorption line ($E_2$), and 3 keV absorption line ($E_3$). The distributions fitted by $\chi^2$-distribution with two dofs (red solid lines). The arrow marks the measured $\Delta C$ from the true data.}
      \label{fig:coad_P}
\end{figure}

As shown in  Fig.~\ref{fig:line_coad}, an excess near 1 keV indicates a clear emission line, while strong evidence of an absorption line near 3 keV is also observed. Additionally, a deviation near 1.6 keV suggests the need for a broad absorption line component.  Significant residuals near 2.2 keV correspond to an absorption edge (2.0--2.4 keV) due to the gold M-shell reflectivity of the detector.  The overall model includes four line components, achieving a fit with $\chi^{2} = 288$ for 258 dofs. For the emission line, the best-fit parameters are a centroid energy of $E_1=1.035\pm{0.011}$ keV, a width of $0.05\pm{0.01}$ keV, and a normalization of $0.32\pm{0.04}~\mathrm{photons~cm^{-2}~s^{-1} }$.  For the broad absorption line, the parameters are $E_2=1.64\pm{0.01}$ keV, a width of $0.17\pm{0.03}$ keV,  and a normalization of  $0.06\pm{0.01}~\mathrm{photons~cm^{-2}~s^{-1} }$. The second absorption line at  $E_3=2.93\pm{0.01}$ keV has a width of $ 0.07\pm{0.02}$ keV and a normalization of $0.065\pm{0.011} ~\mathrm{photons~cm^{-2}~s^{-1} }$. In Fig.~\ref{fig:coad_P}, we show the distributions and fitting result of the three spectral lines of the coadded spectrum. We found the significances of the emission line, $\sigma_{\rm E1} = 14.2\sigma$, and the absorption lines, $\sigma_{\rm E2} = 10.8\sigma$ and $\sigma_{\rm E3} = 11.7\sigma$, respectively. 

\begin{figure}
    \centering
        \includegraphics[width=9cm]{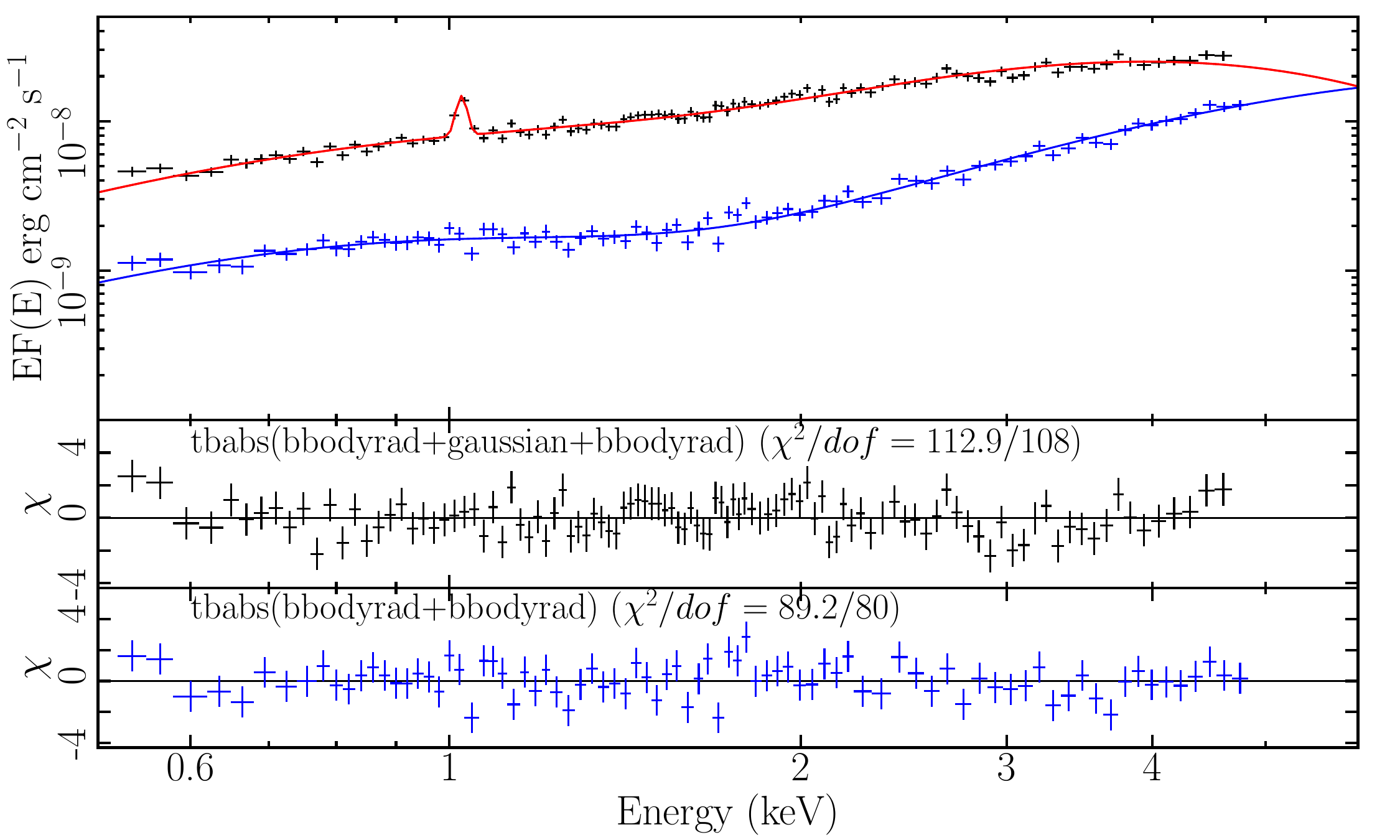}
\caption{Coadded spectroscopy of all burst at the moderate expansion phase (black) and the touchdown moment (blue).  The touchdown spectra show no evidence of line features and the moderate expansion phase shows a narrow 1 keV emission line.}
      \label{fig:line_coad2}
\end{figure}

From the time-resolved spectroscopy reported in \citet{Yuwh2024A&A},  a moderate expansion phase follows the strong expansion phase, with radii that are closely matched. We extracted spectra from the moderate expansion phase for a 0.2 s interval, we coadded these spectra and fitted the same continuum model as for the PRE phases. Additionally, spectra near the touchdown moment were also coadded and analyzed. Residuals from the fits for the moderate expansion phase (in black, middle panel) and the touchdown phase (in blue, bottom panel) are shown in Fig.~\ref{fig:line_coad2}. No line features were detected in the touchdown spectra, but a narrow emission line at $E_{\rm 1}\sim1.024$ keV  was identified in the moderate expansion phase ($R_{\rm bb} = 47\pm{15}$ km), with a significance of $5.0\sigma$ and an improvement in $\chi^{2}$ of 30.

\section{Discussion}
\label{Sec:discussion}
In this work, we have analyzed the spectra of NICER observations between 2017 to present from 4U 1820--30, in which presented strong evidence to support the existence of narrow spectral lines during the PRE phases. By using the Monte Carlo simulations from \citet{Protassov_2002} to estimate the significance, we identified the 1 keV line from 14 bursts and the 3 keV line from 12 burst (see Table~\ref{table:burst_PRE1}). We coadded the spectra of the maximum radius phase and identified three spectral lines in the coadded spectra, namely: the emission line at 1.034 keV and the absorption lines at 1.56 and 2.929 keV, respectively. The most prominent of spectral features is the emission line and the significance of coadded spectrum around $9.4\sigma$. We can infer the possible origin and physically motivated based on the fit results.

\subsection{Plausible origin from the accretion disk}
Spectral lines have recently been observed in several NS LMXBs, particularly during X-ray bursts. In a series of X-ray bursts, a broad Fe-K$\alpha$ emission line was detected with RXTE\ and attributed to reflection from the accretion disk \citep{Keek_2014,int2010,Degenaar18,Salvo2022}. A strong 1 keV emission lines was  observed in SAX J1808.4--3658 \citep{int2013A&A, Bult2019ApJ}, IGR J17062--6143 \citep{Degenaar2013, keek2017, bult2021}, and 4U 1850--087 \citep{lu2024ApJ}. These authors also interpreted the lines as having been caused by the reflection of Fe-K$\alpha$ from accretion disk. 

4U 1820--30 also exhibited clear evidence of spectral line originating from the accretion disk. A prominent absorption line from 4U 1820--30 was observed by NICER after a superburst, with the line exhibiting a shifting from 4.16 to 3.60 keV during the recovery phase of persistent emission. This absorption feature was likely produced by the hydrogen-like Ar K$\alpha$ transition, originating from the inner accretion disk \citep{Peng2024}.

The interpretation of the reflection's origin   presents several challenges. First, no emission or absorption lines were observed during burst $\#\ 2$ (as shown in Fig.~\ref{fig:line}),  despite the high reflection intensity reported during this burst \citep{Yuwh2024A&A}. Notably, the fraction of bolometric flux from disk reflection in this burst is consistent with those bursts where spectral lines were observed. Additionally, all spectral lines were detected exclusively during the strong PRE phase, which is significantly shorter than the duration of reflection.
A second issue involves the significant and systematic spectral shifts in line energies associated with different PRE strengths. \citet{Strohmayer2019} found that the lines in stronger PRE bursts were systematically blueshifted by a factor of 1.046, compared to those observed in weaker bursts. Similarly, from the coadded spectra, we observed the 1 keV line in the maximum radius phase ($E_{1} = 1.039$ keV at $R_{\rm bb}\sim 120 $ km) exhibiting weaker redshifts than in the moderate expansion phase ($E_{1} = 1.024$ keV at $R_{\rm bb}\sim 50 $ km). Therefore, we conclude that the lines originating from the accretion disk might not fully account for the spectral features observed in our data.

\begin{figure}
\includegraphics[width=\hsize]{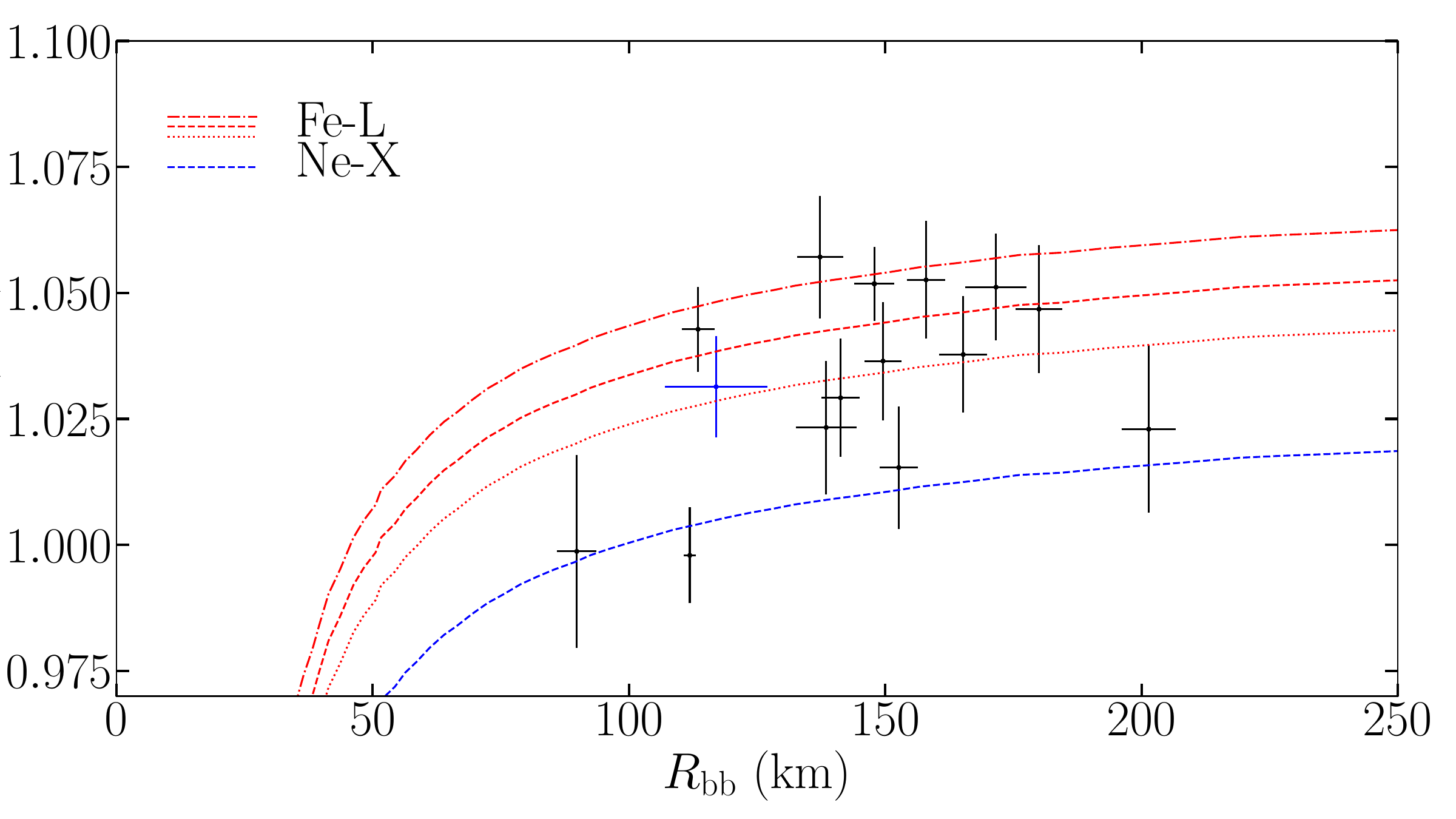}

\caption{1 keV emission lines of 13 bursts with the blackbody radius. The Ly$\alpha$ transition of Ne X in the rest frame at 1.022 keV is plotted with blue dot line. The Fe L shell transitions at 1.045, 1.055 and 1.066 keV are plotted with red dot line, see e.g., \citet{Guichandut2021ApJ}.  }
      \label{fig:line1_e}
\end{figure}

\subsection{Originating from the PRE winds}
Substantial efforts have been made by numerous research groups over the past decades to search for a spectral feature in NS surface. \citet{Cottam02} reported the spectral lines from EXO 0748--676 during its X-ray bursts; however, the detection was quite controversial \citep[see e.g.,][]{Galloway21,Ascenzi2024}. \citet{bult2021} also detected a highly significant absorption line at 3.4 keV in a burst tail from IGR J17062--6143 observed with NICER. They claimed this line was the burning ashes on the stellar surface, such as Ca and Ti.

In UCXB 4U 1820-30, we detected a highly significant absorption line at 3 keV from 12 bursts and a narrow absorption feature at 1.6 keV in one burst. These features are plausibly associated with heavy elements in the burning ashes produced during the bursts.   Possible interpretations of the 3 and 1.6 keV absorption lines include the He-like lines of S XV (2.8839, 3.0325, and 3.1013 keV; \citealt{van2018}), as well as transitions of Fe (XXIII, XXIV, and XXV), Mg XI, or Cr XXIV of $\sim1.6$ keV, respectively \citep{Strohmayer2019}. In contrast to the transient PRE-phase lines, \citet{2023Marino} found persistent features likely originate in the accretion disk, as evidenced by their stability and energy range.

The emission lines detected in 4U 1820--30 could also originate from burning ashes in the photosphere. Two potential explanations include  the Ly$\alpha$ transition of Ne X at 1.022 keV \citep{ne_1995ApJ,keek2017, bult2021} and Fe L-shell transitions, which are produced by irradiation of relatively cool gas \citep{Fe_2008A&A,Degenaar2013,Fe_2024arXiv}. Since 4U 1820--30  accretes helium-rich material  \citep{Stella1987, Rappaport1987,Costantini2012},  the detection of abundant Ne is plausible, since neon is a byproduct of helium burning \citep{ne_1995ApJ}. Additionally,  observations of Fe L-shell lines would be consistent with \textit{XMM-Newton} findings from \citet{Costantini2012}, which reported Fe L absorption features in this system.

To summarize, it is reasonable to suggest that all detected lines originate within the photosphere. Strong PRE bursts likely drive super-Eddington winds, as hydrodynamic simulations indicate that heavy elements can be ejected into the wind, imprinting spectral lines and edges on the burst spectra \citep{Yu2018}. During the expansion phase, the photospheric radius can exceed 100 km, matching the wind solution of \citet{Guichandut2021ApJ}. In our observations, the maximum radius during the PRE bursts exceeds this solution (see Fig.~\ref{fig:line1_e}). In these burst-driven winds, line energies are influenced by gravitational redshift and Doppler blueshift \citep{Strohmayer2019}.  Stronger PRE bursts, reaching larger radii, generate faster outflows and higher blueshifts, with observed blueshifts systematically around 1.046.

However, \citet{Guichandut2021ApJ} found that in steady-state radiation-driven winds within a steady-state model of radiation-driven winds, the wind velocity stabilizes at $v_{\rm max}\sim 0.01c$.  This results in shifts at the photosphere that are predominantly governed by gravitational redshift, with combined effects leading to shifts of less than $\Delta E/E\approx0.02$.  Observed energies slightly deviate from the rest frame under such conditions.  For example, we detected emission lines at 1.03 keV, corresponding to 1.05 keV in the rest frame. If these emission lines have a consistent origin across different bursts, they are more likely to be Fe L-shell transitions with minimal gravitational redshift (see Fig.~\ref{fig:line1_e}). Most NICER bursts align with the predicted combined shifts for Fe L-shell transitions. %

\subsection{The ratio of $E_3/E_1$}

A 1 keV emission line was also observed in the persistent emission of IGR J17062--6143. The high-resolution spectroscopy results led to the interpretation that that this feature was a superposition of several more narrow lines \citep{Degenaar2017,van2018}. \citet{bult2021} speculated that the line measured in burst is also a superposition of several more narrow lines.

\begin{figure}
\includegraphics[width=\hsize]{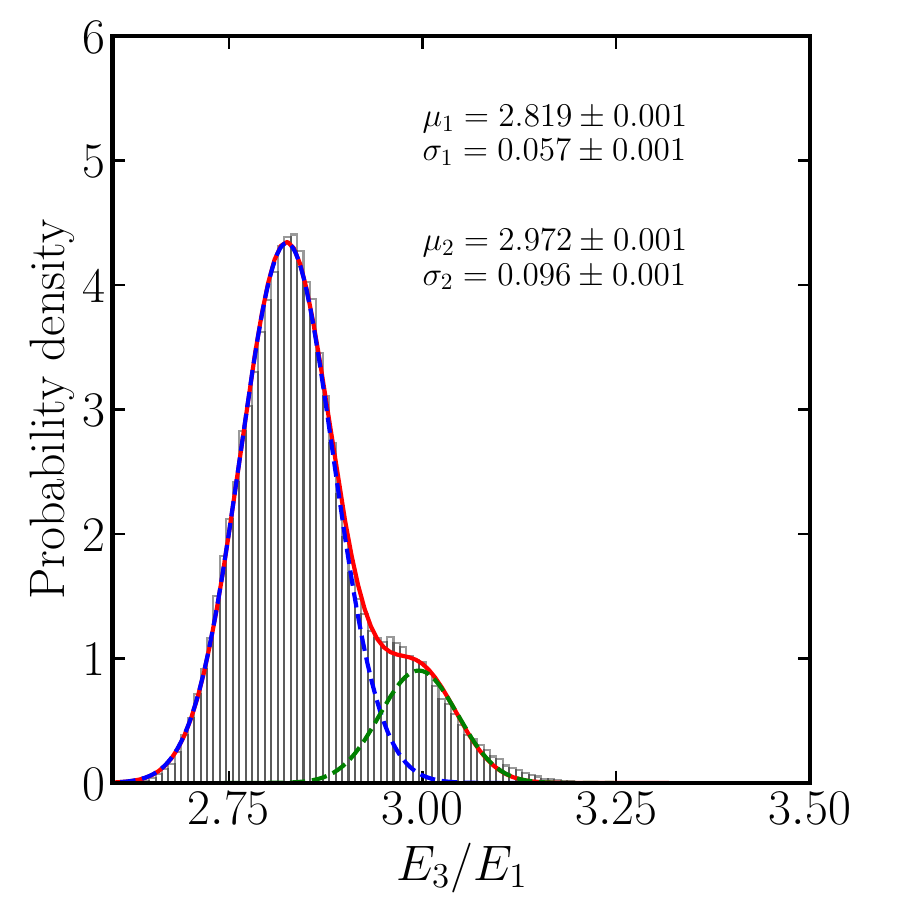}

\caption{Ratios of the centroid energy between $E_{3}$ and $E_{1}$ at the same time. The distributions are well fitted by two Gaussian function. The main peak at 2.819, and secondary at 2.972. }
      \label{fig:line_ratio}
\end{figure}

Assuming that the spectral lines $E_1$ and $E_3$ were both produced by the PRE wind at the same location, the absorption and emission lines in a single spectrum would experience identical gravitational redshift and Doppler blueshift. Consequently, the observed line ratio $E_3/E_1$ would match the ratio measured in the rest frame. Using the measured values of $E_1$ and $E_3$ (see Table~\ref{table:burst_PRE1}), we generated Gaussian distributions for each line based on their centroid energy and width. Monte Carlo simulations were then employed to calculate the line ratio and its uncertainty. Figure~\ref{fig:line_ratio} shows the resulting line ratio ($E_3/E_1$) distributions, which are well-fitted by two Gaussian functions with $\mu_1 = 2.819$ and $\sigma_1 = 0.057$ for the main peak, and $\mu_2 = 2.972$ and $\sigma_2 = 0.096$ for the second peak. If $E_3$ represents a single line, this suggests that the emission line ($E_1$) is a superposition of several narrower lines rather than a single broad feature. Based on the mean values of the two peaks, we infer an energy ratio of approximately 1.05 for two narrow lines around 1 keV.

\subsection{The future prospective}
At present, the direct and precious identification of spectral features is challenging due to the moderate energy resolution of NICER detectors. Previous attempts to detect spectral features during X-ray bursts using \xmm\ and {\it Chandra}, such as in EXO 0748--676 and Rapid Burster \citep{Zand2017, cottam2008ApJ}, involved stacking numerous burst spectra. However, no spectral lines were detected in those attempts. This lack of detection might be due to energy shifts of spectral lines during bursts or broadening effects caused by the rapid rotation of NSs. Several upcoming missions with high energy resolution and large collection areas, such as the Hot Universe Baryon Surveyor (HUBS) and the Advanced Telescope for High Energy Astrophysics  \citep[{\it NewAthena};][]{newath2024,newAth2025}, aim to overcome these limitations with high energy resolution and large collection areas.  In particular, HUBS features an X-ray spectrometer capable of achieving eV-level precision in the 
0.1--2.0 keV band \citep{2020hubs, 2022hubs, 2024hubs}.  It also offers a large field of view and an effective area exceeding $> 500 ~{\rm cm^{2}}$ at 1 keV \citep{hubs_arf2020}. These capabilities make HUBS highly optimized for detecting emission or absorption lines.

\begin{figure}

    \centering
        \includegraphics[width=\hsize]{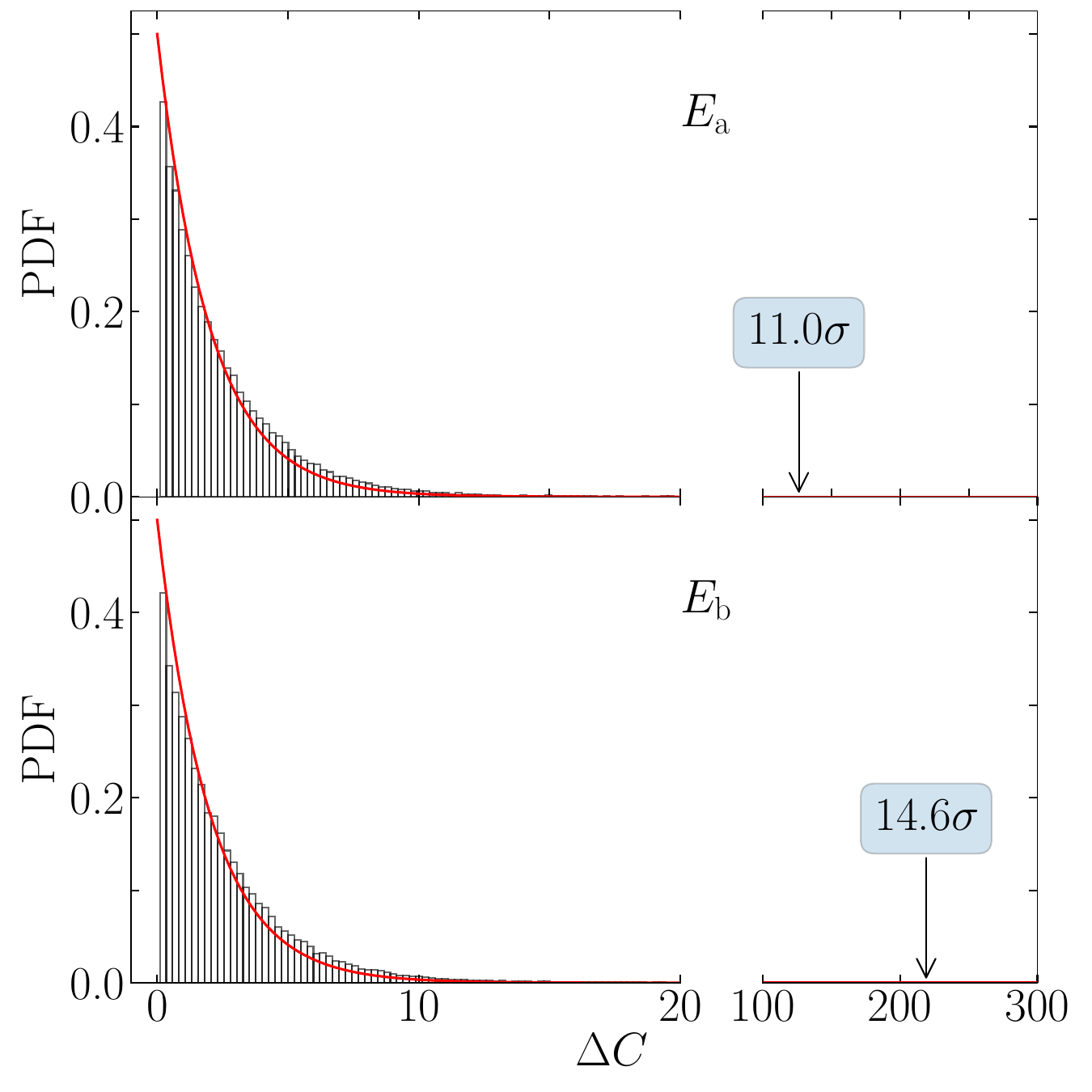}

\caption{Distributions of the $\Delta C$ for $10^5$  simulations for the lines of the mock HUBS spectra. From top to bottom:\ Distributions of the  simulations for 1.005 ($E_{\rm a}$) and 1.06 keV ($E_{\rm b}$) emission lines. The distributions fitted by $\chi^2$-distribution with two dofs (red dashed-dotted lines). The arrow marks the measured $\Delta C$ from the true data.}
      \label{fig:hubs_P}
\end{figure}

4U 1820--30 is an excellent target for HUBS observations, given its relatively low Galactic column density and energetic X-ray bursts in the soft X-ray band. To evaluate HUBS’s ability to resolve narrow lines near 1 keV, we generated a mock spectrum using the fitted parameters of burst \#8 and the current ARF and RMF of  HUBS. Using the XSPEC tool \texttt{fakeit}, we simulated spectra with a one-second exposure time, incorporating two emission lines at 1.005 and 1.06 keV  (both with width of 0.01 keV).  The spectra were grouped with minimum counts of 20. The simulated spectrum clearly resolved the two spectral features, which were successfully fitted using a two-Gaussian model. The fitted line energies were $E_a = 1.005\pm{0.002} \, \text{keV}$ and $E_b = 1.061\pm{0.002} \, \text{keV}$.  The MC simulations estimated significances of approximately $\sim 10.4\sigma$  for $E_{a}$ and $\sim 12.3\sigma$ for $E_{b}$ (see Fig.~\ref{fig:hubs_P}). By combining these measured energy shifts with the effects of rotational broadening, the NS mass and radius can be accurately determined using a model-independent approach. This will provide crucial constraints on the equation of state for compact objects, offering valuable insights into their fundamental properties \citep[see e.g.,][]{Chang06,2008jonker,li2018ApJ}.

\begin{figure}
\includegraphics[width=\hsize]{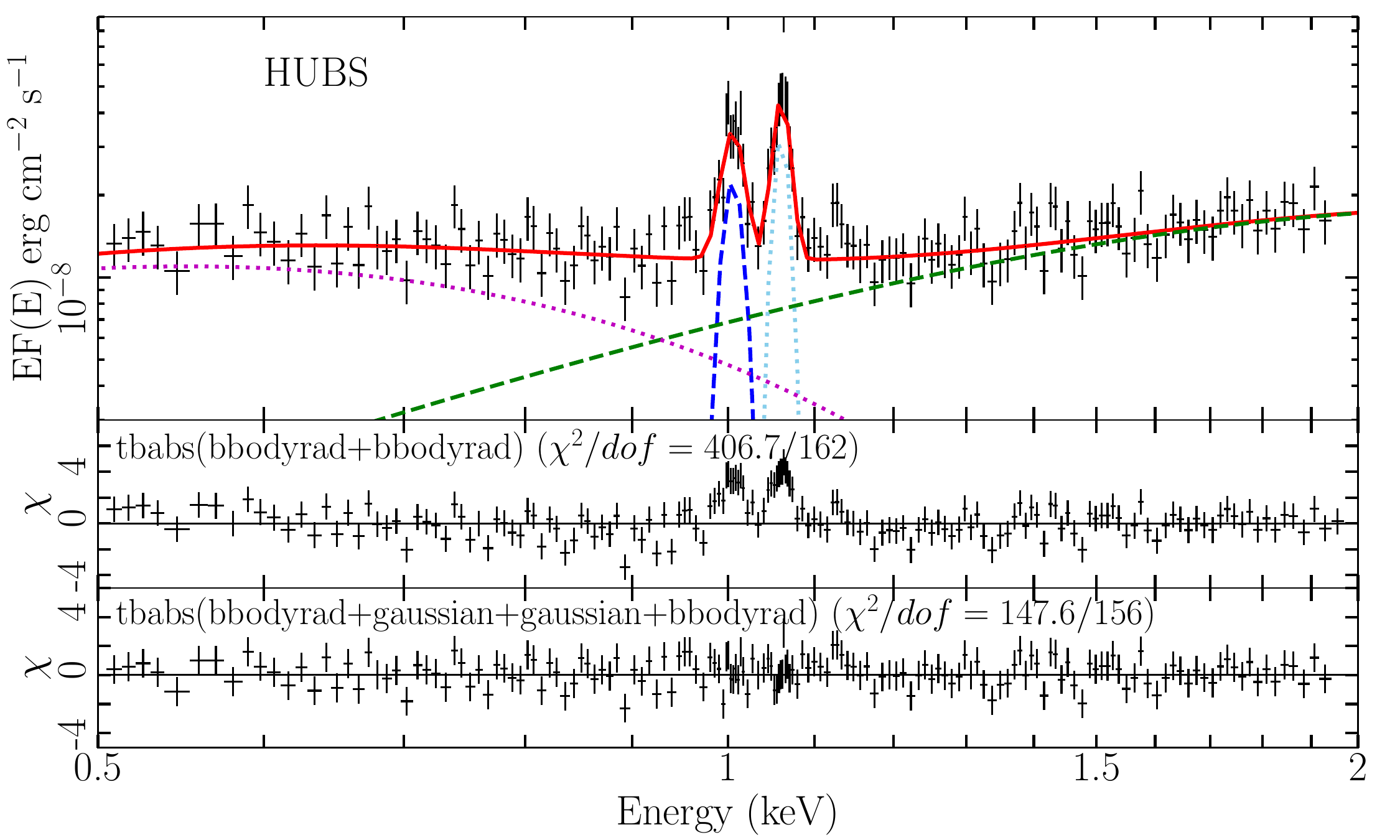}

\caption{Mock HUBS X-ray spectrum with the continuum parameters from burst \#8 and two additional narrow emission lines. The best fitting model is ${\tt Tbabs} \times ({\tt bbodyrad + gaussian +gaussian+bbodyrad })$. The dashed lines represent the two Gaussian lines. The best-fit model is plotted as red solid line. The burst blackbody and disk blackbody components are shown in pink and green dotted lines, respectively.}
      \label{fig:hubs}
\end{figure}

When the current work was in progress, \citet{Barra2025A&A} was also analyzing the spectral lines in 12 type I X-ray bursts from the same NICER observations.  Compared with their work, our data set include three more X-ray bursts from the UFA event files (see Table~\ref{table:data}). \citet{Barra2025A&A} applied different models to  the burst continuum spectra and also identified these three lines.  Moreover, \citet{Barra2025A&A} obtained other emission residuals at 2 keV (Si XIV) and 2.6 keV (S XVI) using state-of-art plasma codes available within SPEX with a phenomenological continuum.  Generally, our results are broadly consistent with \citet{Barra2025A&A}. 

\section{Conclusions}
\label{Sec:conclusion}
In this study, we carried out a comprehensive spectral analysis of 4U 1820--30 using NICER observations, focusing on the search for emission and absorption lines in 15 PRE bursts.  We identified a 1 keV emission line from 14 bursts, a 1.6 keV absorption line from one burst, and a 3 keV absorption line from 12 bursts. The main results are generally consistent with those derived from the first five bursts presented by \citet{Strohmayer2019}. However, our findings indicate a lower confidence level for the 1.6 keV absorption line.  These lines were detected in coadded spectra, revealing a significant and systematic shift in their energies between bursts with varying PRE radii. These shifts are consistent with predictions from burst-driven wind models, suggesting that the spectral lines originate from heavy elements in the PRE winds. Based on the predicted line energies in winds, we attribute the 1 keV emission line to Fe L-shell transitions. Furthermore, our analysis indicates that the 1 keV line is a superposition of several narrow lines rather than a single broad feature. Simulated mock observations with a one-second exposure time demonstrate that HUBS data can effectively resolve the 1 keV emission lines during bursts, offering new opportunities for precise spectral studies.

\begin{acknowledgements}
We appreciate the referee for the valuable comments and suggestions, which improves the manuscript. This work was supported by the Major Science and Technology Program of Xinjiang Uygur Autonomous Region (No. 2022A03013-3). Z.S.L. and Y.Y.P. were supported by National Natural Science Foundation of China (12103042, 12273030, 12122302, 12173103, U1938107). This work made use of data from the High Energy Astrophysics Science Archive Research Center (HEASARC), provided by NASA’s Goddard Space Flight Center. 

\end{acknowledgements}

\bibliography{file}{}
\bibliographystyle{aa}

\begin{appendix} %

\begin{table*}[h!]
\section{Table} \label{Sec:App_Table}

\caption{Burst observations. }
\label{table:data}
\begin{center} 
 
 \resizebox{\linewidth}{!}{\begin{tabular}{ccccccccccc} 
 \hline
 \hline\\ %
 {\centering  Burst No. } &
 {\centering  Obs-ID} &
 {\centering  Date} &
 {\centering  Time} &
 {\centering  MJD  } &
 {\centering  Peak Rate\tablefootmark{a}} &
 {\centering  $\Delta t$\tablefootmark{b} } &
 \\
 $\#$& & (YYYY-MM-DD) & (UT)& (TT) &($\mathrm{10^{4} ~ c ~ s^{-1}}$)& (s) & \\ [0.01cm] \hline
 1  & 1050300108                 &2017-08-29&08:58:27&57994.37393 &2.15 &  0.3-0.8 &  \\
 2  & 1050300108                 &2017-08-29&11:08:51&57994.46448 &1.56 &  0.4-1.4 &  \\
 3  & 1050300109                 &2017-08-30&05:24:49&57995.22557 &2.28 &  0.4-0.8 &  \\
 4  & 1050300109                 &2017-08-30&08:11:55&57995.34162 &2.08 &  0.4-0.8 &  \\
 5  & 1050300109                 &2017-08-30&14:32:38&57995.60600 &1.89 &  0.9-1.4 &  \\
 6  & 2050300104                 &2019-06-07&07:05:51&58641.29573 &2.52 &  0.5-1.1 &  \\
 7  & 2050300108                 &2019-06-12&20:09:49&58646.84015 &1.96 &  0.1-0.8 &  \\
 8  & 2050300110                 &2019-06-14&07:50:25&58648.32669 &2.25 &  0.4-1.4 &  \\
 9  & 2050300115\tablefootmark{c}&2019-06-21&02:12:31&58655.09203 &2.59 &  0.2-0.7 &  \\
 10 & 2050300119                 &2019-06-26&07:31:06&58660.31327 &2.60 &  0.3-1.0 &  \\
 11 & 2050300119\tablefootmark{d}&2019-06-26&18:30:54&58660.77147 &2.58 &  0.4-1.1 &  \\
 12 & 2050300120\tablefootmark{c}&2019-06-27&18:52:02&58661.78614 &2.22 &  0.8-1.4 &  \\
 13 & 2050300122\tablefootmark{c}&2019-06-29&23:37:30&58663.98438 &2.57 &  0.4-1.0 &  \\
 14 & 2050300124                 &2019-07-01&14:15:25&58665.59405 &2.54 &  0.4-1.1 &  \\
 15 & 4680010101                 &2021-05-02&14:41:42&59336.61230 &2.57 &  0.6-1.2 &  \\
 \hline 
 \end{tabular} }
 \end{center}
 \tablefoot{
 \tablefoottext{a}{The peak rates with persistent emission subtracted are measured from the 0.1 s light curves in the energy range of 0.5--10 keV.}\\
 \tablefoottext{b}{The start/end times (relative to burst onset) used to extract PRE phase spectra.}\\
 \tablefoottext{c} {The X-ray bursts in the ufa event file.}\\
 \tablefoottext{d}{ The tail of burst \#10 was truncated due to data gap.}\\
 }
 \end{table*}

\end{appendix}

\end{document}